# Limits on Three-Point Correlations in the $COBE^1$ DMR First Year Anisotropy Maps

G. Hinshaw[2,3], A. Kogut[2], K.M. Górski[4,5], A.J. Banday[4], C.L. Bennett[6], C. Lineweaver[7], P. Lubin[8], G.F. Smoot[7] & E.L. Wright[9]

[1]The National Aeronautics and Space Administration/Goddard Space Flight Center (NASA/GSFC) is responsible for the design, development, and operation of the Cosmic Background Explorer (*COBE*). Scientific guidance is provided by the *COBE* Science Working Group. GSFC is also responsible for the development of the analysis software and for the production of the mission data sets.

[2]Hughes STX Corporation, Code 685.9, NASA/GSFC, Greenbelt MD 20771.

[3]e-mail: hinshaw@stkitt.gsfc.nasa.gov

[4]Universities Space Research Assoc., Code 685.9, NASA/GSFC, Greenbelt MD 20771.

[5]On leave from Warsaw University Observatory, Poland.

[6]NASA Goddard Space Flight Center, Code 685, Greenbelt MD 20771.

[7]LBL, SSL, & CfPA, Bldg 50-351, University of California, Berkeley CA 94720.

[8]UCSB Physics Department, Santa Barbara CA 93106.

[9]UCLA Astronomy Department, Los Angeles CA 90024-1562.

## ABSTRACT

We compute the three-point temperature correlation function of the *COBE* Differential Microwave Radiometer (DMR) first-year sky maps to search for non-Gaussian temperature fluctuations. The level of fluctuations seen in the computed correlation function are too large to be attributable solely to instrument noise. However the fluctuations are consistent with the level expected to result from a superposition of instrument noise and sky signal arising from a Gaussian power law model of initial fluctuations, with a quadrupole normalized amplitude of 17 $\mu$K and a power law spectral index $n = 1$. We place limits on the amplitude of intrinsic three-point correlations with a variety of predicted functional forms.

*Subject headings:* cosmic microwave background — cosmology: observations

## 1. Introduction

NASA's *COBE* collaboration has reported the detection of anisotropies in the cosmic microwave background (CMB) radiation (Smoot et al. 1992; Bennett et al. 1992a; Wright et al. 1992 ). The amplitude of these anisotropies is substantially greater than the limits on possible systematic effects remaining in the data (Kogut et al. 1992), while the pattern of anisotropy is not correlated with any known Galactic or extragalactic foreground (Bennett et al. 1992a; Bennett et al. 1993). A useful characterization of CMB anisotropies is provided by the $n$-point correlation functions, the average product of temperatures evaluated at $n$ points in the sky with fixed relative orientation. The *COBE* collaboration reported a positive detection of the two-point temperature correlation function (Smoot et al. 1992; Wright et al. 1992), and found it to be consistent with inflationary and other cosmologies. One of the basic predictions of inflationary models is that they produce a nearly scale-invariant power spectrum of primordial density fluctuations with a Gaussian distribution and random phases. It follows that large angular scale CMB temperature anisotropies should have vanishing three-point correlations on average (though any particular cosmic observer will, in general, measure small non-zero three-point correlations due to cosmic variance). In contrast, there are competing theories of structure formation (eg., theories based on topological defects or late-time phase transitions) that are compatible with present CMB observations for which the distribution of density and temperature fluctuations are non-Gaussian. Such models can give rise to non-vanishing three-point temperature correlations, thus a crucial test of the Gaussian nature of the primordial density fluctuations is provided by this function. Smoot et al. (1994) discuss a number of additional tests that can be applied to the DMR sky maps to search for evidence of non-Gaussian fluctuations.

In this paper we compute the three-point correlation function of the first-year DMR sky maps and compare the results to detailed Monte Carlo simulations of Gaussian, scale-invariant power law models. We conclude that there is evidence for non-vanishing three-point correlations in the data, but at a level consistent with the uncertainty associated with cosmic variance. In the absence of a positive detection of intrinsic three-point correlations we can place upper limits on correlations with selected, theoretically motivated functional forms. Falk, Rangarajan, & Srednicki (1993) have considered the effect of a cubic term in the effective potential of an inflationary field $\varphi$ and have concluded that the leading contribution to the three-point function can be expressed as a quadratic function of the corresponding two-point function. Luo & Schramm (1993) have argued that nonlinear evolution of the gravitational potential $\phi$ will give rise to a similar form for the three-point function with an amplitude that likely dominates the inflationary contribution.

In addition, Luo & Schramm (1993) suggest that a late-time phase transition could produce a contribution that may be expressed as a cubic function of the corresponding two-point function. We place constraints on the amplitudes of such contributions.

The DMR experiment (Smoot et al. 1990; Bennett et al. 1992b) has produced two independent microwave maps (A and B) at each of 3 frequencies (31.5, 53 and 90 GHz). The results presented here are based on the first year of data, and we consider only the relatively sensitive 53 and 90 GHz maps. In all cases we restrict our analysis to Galactic latitudes $|b| > 20°$, and we subtract a best-fit monopole and dipole from the data. The general three-point correlation function is the average product of three temperatures with a fixed relative separation on the sky: $C_3(\theta_1, \theta_2, \theta_3) = \langle T(\hat{n}_1)T(\hat{n}_2)T(\hat{n}_3) \rangle$ where $\hat{n}_1 \cdot \hat{n}_2 = \cos\theta_1$, $\hat{n}_2 \cdot \hat{n}_3 = \cos\theta_2$ and $\hat{n}_3 \cdot \hat{n}_1 = \cos\theta_3$. Due to computational constraints we evaluate only two special cases: 1) the direction vectors $\hat{n}_i$, $\hat{n}_j$, and $\hat{n}_k$ form an equilateral triangle, $\theta_1 = \theta_2 = \theta_3$ (the equilateral case), and 2) $\hat{n}_i$ is nearly parallel to $\hat{n}_j$ (the pseudo-collapsed case). We motivate and define the use of "nearly" in §3 below. Our notation is as follows: latin indices $i$, $j$, and $k$ refer to sky pixels, while greek indices $\alpha$ and $\beta$ refer to bins of angular separation, which we take to be of width 2.6 degrees, the width of our sky pixels. Additionally, the use of $\theta$ as an argument to a correlation function denotes a value of angular separation in degrees.

## 2. The Equilateral Case

The DMR sky maps are binned into 6144 nearly equal-area pixels, 4016 of which pass the 20 degree Galaxy cut. We define the equilateral three-point correlation function to be

$$C_3^{(e)} = \sum_{i,j,k} w_i w_j w_k \, T_i T_j T_k / \sum_{i,j,k} w_i w_j w_k$$

where the sum is restricted to pixel triples $(i, j, k)$ for which all three pixel separations reside in a single angular separation bin, $T_i$ is the observed temperature in pixel $i$ after monopole and dipole subtraction, and $w_i$ is the statistical weight of pixel $i$: $w_i = 1/\sigma_i^2 = N_i/\sigma_0^2$, where $N_i$ is the number of observations of pixel $i$, and $\sigma_0$ is the rms temperature fluctuation per observation due to instrument noise.

We have evaluated this statistic for the 53 and 90 GHz DMR maps and present the results in Figure 1. The left-hand panels of the figure show results for both the (A+B)/2 ("sum") maps, which include both sky signal and instrument noise, and the (A−B)/2 ("difference") maps which contain only instrument noise. The grey band in these panels represents the rms scatter computed from 2000 Monte Carlo realizations of the instrument noise. The right-hand panels repeat the results for the sum maps, but in these panels the grey bands represent the rms scatter computed from 2000 Monte Carlo realizations of simulated CMB sky signal plus instrument noise. In the present analysis we have simulated the CMB signal using a Gaussian, scale-invariant power law model with random phases and a mean quadrupole normalization of 17 $\mu$K. Power from $l = 2$ to $l = 40$ filtered through the DMR window function (Wright et al. 1994) is included. Instrument noise was simulated for each channel A and B (with the appropriate DMR sky coverage pattern) using the following values for the instrument noise per 0.5 second observation: $\sigma_0 = 24.2$, 28.2, 47.8, and 35.6 mK (thermodynamic temperature) for channel 53A, 53B, 90A, and 90B respectively. For each realization of sky signal and noise, we form sum and difference maps and subtract best-fit (for $|b| > 20°$) monopoles and dipoles from each. From the ensemble of realizations we compute the rms scatter in each of the 48 angular separations bins, as well as the full covariance matrix to account for bin-to-bin correlations induced by the sky signal.

We have run a set of noiseless Monte Carlo simulations to compare the results of our routine with existing analytic predictions for the level of cosmic variance expected in the three-point function (Srednicki 1993). First we have generated a set of 400 Harrison-Zel'dovich ($n = 1$) skies with unit quadrupole normalization including power from $l = 3$ to $l = 40$, filtered through the DMR window function. For each realization in this ensemble we compute the three-point function using uniform weights over the entire sky. The rms scatter of these functions may then be directly compared to Figure 2 of Srednicki (1993); our results agree to within a few percent. The Galactic plane cut is an additional source of theoretical uncertainty (a form of "sample variance"). We find that restricting the sky coverage to $|b| > 20°$ produces an rms that is larger than the full-sky case by a factor of $\sim 1.5$, somewhat larger than the factor of 1.23 predicted by Srednicki, based on the work of Scott, Srednicki, & White (1993). An additional source of sample variance in our data arises from the pattern of weights used to estimate the three-point function. Including the quadrupole in our noise-free simulations, and weighting the computation of the three-point function by the inverse of the DMR noise variance, we find the rms to be a factor of $\sim 5 - 8$ times larger (depending on the angular separation) than the uniformly weighted, full-sky, no-quadrupole case. It is advantageous to weight the computation of the three-point function to minimize the noise variance since the signal-to-noise ratio in our first-year maps is relatively poor. However, with additional years of data, cosmic variance will dominate noise uncertainties and it will be preferable to weight the pixels uniformly to minimize this source of uncertainty.

The hypothesis that the computed three-point function is consistent with zero is tested by computing $\chi^2 = \sum_{\alpha,\beta} C_\alpha \left(M^{-1}\right)_{\alpha\beta} C_\beta$ where $C_\alpha$ is the observed three-point function in angular separation bin $\alpha$, and $M_{\alpha\beta}$ is the covariance matrix computed from the simulated three-point functions. The observed values of $\chi^2$ are given in Table 1, the numbers given in parentheses are the percentage of simulations for which $\chi^2$ exceeded the observed value. Consider first the hypothesis that the observed three-point functions are consistent with *no* sky signal, ie., that the observed fluctuations are due only to instrument noise. In this case we evaluate $\chi^2$ using the three-point functions observed in the sum maps and the covariance matrix derived from the ensemble of difference maps which have the same noise properties as the sum maps. We find $\chi^2 = 69$ and 57 for the 53 and 90 GHz data respectively; only 4% of our 53 GHz difference simulations and 18% of our 90 GHz difference simulations had higher values of $\chi^2$. We conclude that there exist non-vanishing equilateral three-point correlations in the observed CMB sky at the ~95% confidence level. We next test the hypothesis that the fluctuations seen in the three-point function are consistent with the level expected to arise from a superposition of instrument noise and sky signal arising from Gaussian, scale-invariant power law fluctuations. In this case $\chi^2$ is evaluated using the covariance matrix computed from the ensemble of sum maps; we find $\chi^2 = 49$ and 50 for the 53 and 90 GHz data respectively. As indicated in Table 1, these values are well within the range seen in the simulations, which indicates that the fluctuations observed in the equilateral three-point function are consistent with a superposition of instrument noise and Gaussian CMB fluctuations.

As a check on the noise levels in our simulations we have evaluated $\chi^2$ using the three-point functions observed in the difference maps. We find the 53 GHz value to be well within the expected range while the 90 GHz value is not: only 1 of our 2000 simulations had as large a $\chi^2$ for 48 degrees of freedom. We attribute this to a weak anti-correlation between the noise in the 90 GHz A and B channels, which we have recently identified by other analyses. As a consequence the 90 GHz difference map is not a reliable estimator of the instrument noise level in the maps. We expect to correct this anti-correlation in future releases of the data and to discuss it more fully in a future publication. We emphasize that no present or previously reported results are compromised by this effect.

In the absence of a positive detection of intrinsic three-point correlations we can place limits on the amplitudes of contributions with certain predicted functional forms. Falk et al. (1993) and Luo & Schramm (1993) argue that the equilateral configuration of the three-point function will have leading contributions of the form $C_3^{(e)}(\theta) \approx k_2 C_2^2(\theta) + k_3 C_2^3(\theta)$, where $C_2(\theta)$ is the corresponding two-point function, and $k_2$ and $k_3$ are constants. To place limits on $k_2$ and $k_3$ one should ideally invoke a full Monte Carlo simulation of CMB skies with the appropriate non-Gaussian effects included. However, since the above predictions are only approximate, we feel it is sufficient at this time to adopt the theoretical form for the mean of $C_2(\theta)$ as a basis for our model functions, and to perform the fits using the errors obtained from our Gaussian Monte Carlo simulations. In particular we define $C_2(\theta)$ as (Abbott &

Wise 1984)

$$C_2(\theta) = (6/5) \sum_{l=2}^{40} (2l+1)/(l(l+1)) \, G_l^2 \, P_l(\theta)$$

where the $G_l$'s, given by Wright et al. (1994), define the DMR window function, and the $P_l$'s are the Legendre polynomials. For reference, $C_2(0) = 4.24$. We then perform least-squares fits to the data by minimizing a $\chi^2$ defined to be

$$\chi^2 = \sum_{\alpha,\beta} (C_\alpha - k_m C_2^m(\theta)) \left(M^{-1}\right)_{\alpha\beta} (C_\beta - k_m C_2^m(\theta))$$

for $m = 2, 3$. We evaluate the model function $C_2(\theta)$ at the center of each angular separation bin. The covariance matrix is derived from the ensemble of sum maps; thus the quoted confidence intervals include the effects of cosmic variance. Since the two functional forms are nearly degenerate, we cannot place meaningful simultaneous constraints on the two parameters $k_2$ and $k_3$, so we perform each fit separately. We define approximate 68% confidence intervals for $k_2$ and $k_3$ by the condition $\chi^2(k_m) < \chi^2_{min} + 1$ for $m = 2, 3$. The results of these fits are given in Table 2, which also includes weighted averages. Both $k_2$ and $k_3$ are consistent with zero.

## 3. The Pseudo-Collapsed Case

The simplest configuration of the three-point correlation function is that in which two legs of the correlation function are evaluated at a common point in the sky. SubbaRao et al. (1993) have proposed the use of the "collapsed" three-point function as an effective probe of non-Gaussian structure in the CMB. Using the same notation as above, we define the collapsed three-point function to be

$$C_3^{(c)} = \sum_{i,j} w_i w_j^2 \, T_i T_j^2 / \sum_{i,j} w_i w_j^2$$

where the sum is over all pixel-pairs within a given angular separation bin. Unfortunately, because of the quadratic term, $T_j^2$, in this statistic, the collapsed three-point function suffers from a severe noise bias. In other words $C_3^{(c)}$ is not a central estimator of the true collapsed three-point function in the presence of instrument noise. To see this consider an ensemble average of $C_3^{(c)}$ over many realizations of the instrument noise, with a fixed true sky temperature. We can decompose the observed sky temperature $T_i$ into a true temperature $t_i$ and a noise contribution $n_i$: $T_i = t_i + n_i$. Then, by using the facts that $\langle t_i^p \rangle = t_i^p$, $\langle n_i \rangle = \langle n_i^3 \rangle = 0$, and $\langle n_i^2 \rangle = \sigma_i^2$ we obtain

$$\langle C_3^{(c)} \rangle = \sum_{i,j} w_i w_j^2 \, t_i t_j^2 / \sum_{i,j} w_i w_j^2 \; + \; \sum_{i,j} w_i w_j^2 \, t_i \sigma_j^2 / \sum_{i,j} w_i w_j^2$$

The first term is the "true" three-point function of the sky temperature, while the second is a bias term that is a cross-correlation of the true sky temperature with the average noise pattern. Given the current DMR noise levels and the apparent absence of three-point correlations in the CMB anisotropies, the bias term dominates the behavior of this statistic. We have investigated a number of ways around this problem. For example, we could select the weights $w_i$ in such a way as to make the bias term cancel based on the average noise properties of the map. However, detailed Monte Carlo simulations show that no choice of weights simultaneously removes the noise bias while recovering the "true" three-point function. Alternatively we can evaluate the collapsed function with a subset of the maps, wherein we replace the quadratic term $T_j^2$ with the expression $T_{A,j}T_{B,j}$ where $T_A$ is the observed temperature in the channel A map, and likewise for $T_B$. Since the noise in the two channels is uncorrelated (the 90 GHz channels notwithstanding, see above), there is no surviving bias. However one does sacrifice significant sensitivity with this scheme. Smoot et al. (1994) give a more detailed discussion of noise bias in higher-order statistical quantities.

The technique adopted in this paper for overcoming the noise bias is to evaluate a "pseudo-collapsed" three-point correlation function. We define this statistic as

$$C_3^{(pc)} = \sum_{i,j,k} w_i w_j w_k \, T_i T_j T_k / \sum_{i,j,k} w_i w_j w_k$$

where the sum on $j$ is over all pixels that are nearest neighbors to $i$, and the sum on $k$ is over all pixels (except $j$) within a given angular separation bin of $i$. Since the sky map pixels are 2.6° across and our beam has a 7° FWHM, the nearest neighbor pixels will have correlated sky signal but uncorrelated instrument noise. Additionally, since most pixels have eight nearest neighbors, we are summing over roughly four times as many independent pixel triples as in the true collapsed case, which significantly enhances the sensitivity.

We have evaluated this statistic for the 53 and 90 GHz DMR maps and present the results in Figure 2, which has the same format as Figure 1. As with the equilateral case, we test the hypothesis that the data are consistent with vanishing three-point correlations by computing $\chi^2 = \sum_{\alpha,\beta} C_\alpha \, (M^{-1})_{\alpha\beta} \, C_\beta$ where $C_\alpha$ is now the observed pseudo-collapsed three-point function in angular separation bin $\alpha$, and $M_{\alpha\beta}$ is the covariance matrix computed from the ensemble of simulated, pseudo-collapsed three-point functions. The values of $\chi^2$ are given in Table 3, which has the same format as Table 1. The $\chi^2$ values for the hypothesis that the three-point correlations in the sum maps are consistent with instrument noise are 138 and 127 for the 53 and 90 GHz data respectively. Only 0.1% of our 53 and 90 GHz difference simulations had a larger $\chi^2$. We conclude that there exist non-vanishing pseudo-collapsed three-point correlations in our sky at the ∼99.9% confidence level. However, when we include the effects of cosmic variance, the $\chi^2$ values diminish considerably. Both the 53 and 90 GHz values fall well within the range of our simulations, as indicated in Table 3. Thus cosmic variance can comfortably explain the observed level of pseudo-collapsed three-point correlations.

The functional form predicted by Luo & Schramm (1993) for the pseudo-collapsed configuration of the three-point function is

$$C_3^{(pc)}(\theta) = k_2[2C_2(2.6°)C_2(\theta) + C_2^2(\theta)] + k_3 C_2(2.6°)C_2^2(\theta)$$

where we have used the fact that the mean separation between nearest neighbor pixels is $\sim 2.6°$. For the form of $C_2$ given in §2 we find $C_2(2.6°) = 4.09$. Using the above model functions and the covariance matrices computed from the ensemble of pseudo-collapsed functions, we perform least squares fits for $k_2$ and $k_3$ as in §2. The results are given in Table 4. As before, the results are consistent with zero; additionally, since the pseudo-collapsed configuration has greater sensitivity, the confidence intervals are somewhat smaller than those derived from the equilateral configuration.

## 4. Conclusions

We have evaluated the three-point temperature correlation function for the first year *COBE* DMR sky maps and find evidence for non-zero three-point correlations in the data. However, we demonstrate that the observed level of fluctuations are consistent with the level expected to result from the superposition of instrument noise and CMB sky signal arising from a Gaussian, scale-invariant power law model of initial fluctuations, with a quadrupole normalized amplitude of 17 $\mu$K. We have placed limits on the amplitudes of intrinsic three-point correlations with specific, theoretically-motivated functional forms.

Given that the three-point function is a cubic statistic, the noise levels will diminish relatively rapidly ($\propto time^{-3/2}$) with additional data. In the four year 53 GHz maps, the rms noise per 2.6° angular separation bin will be roughly eight times smaller than the levels depicted in Figures 1*a* and 2*a*. This corresponds to $\sim(11\text{-}12\ \mu\text{K})^3$ for the equilateral configuration and $\sim(6\text{-}9\ \mu\text{K})^3$ for the pseudo-collapsed configuration, depending on angular separation. Thus, our sensitivity to three-point correlations will ultimately be limited by cosmic variance.

Table 1. $\chi^2$ Values for the Equilateral Configuration[a]

| Data[b] | $\chi^2(M^{(+)}_{\alpha\beta})$ | $\chi^2(M^{(-)}_{\alpha\beta})$[c] |
|---|---|---|
| 53 GHz $C^{(+)}_\alpha$ | 49 (42%) | 69 (4%) |
| 53 GHz $C^{(-)}_\alpha$ | ... | 51 (36%) |
| 90 GHz $C^{(+)}_\alpha$ | 50 (39%) | 57 (18%) |
| 90 GHz $C^{(-)}_\alpha$ | ... | 87 (0.05%) |

[a]There are 48 angular separation bins in the equilateral three-point function.
[b]$C^{(\pm)}_\alpha$ denotes a three-point function computed from an (A±B)/2 map.
[c]$M^{(\pm)}_{\alpha\beta}$ denotes a covariance matrix computed from an ensemble of simulated (A±B)/2 maps; separate covariance matrices were computed for the 53 and 90 GHz noise levels.

Table 2. 68% Confidence Intervals, in $\mu K^3$, for $k_2$ and $k_3$[a]

| Frequency[b] | $k_2$[c] | $k_3$[d] |
|---|---|---|
| 53 GHz | $-324 < k_2 < 2841$ | $-167 < k_3 < 750$ |
| 90 GHz | $-7325 < k_2 < -431$ | $-2020 < k_3 < 118$ |
| Average | $-1074 < k_2 < 1803$ | $-324 < k_3 < 519$ |

[a]Approximate 68% confidence intervals, derived from the equilateral configuration, include uncertainties due to cosmic variance.
[b]The line denoted "Average" gives weighted average confidence intervals under the assumption that the 53 and 90 GHz uncertainties are uncorrelated.
[c]$k_2$ is the coefficient of the model function $C_2^2(\theta)$
[d]$k_3$ is the coefficient of the model function $C_2^3(\theta)$

Table 3. $\chi^2$ Values for the Pseudo-collapsed Configuration[a]

| Data[b] | $\chi^2(M_{\alpha\beta}^{(+)})$ | $\chi^2(M_{\alpha\beta}^{(-)})$[c] |
|---|---|---|
| 53 GHz $C_\alpha^{(+)}$ | 50 (83%) | 138 (0.01%) |
| 53 GHz $C_\alpha^{(-)}$ | ... | 71 (42%) |
| 90 GHz $C_\alpha^{(+)}$ | 89 (18%) | 127 (0.01%) |
| 90 GHz $C_\alpha^{(-)}$ | ... | 114 (0.9%) |

[a]There are 71 angular separation bins in the pseudo-collapsed three-point function.
[b]$C_\alpha^{(\pm)}$ denotes a three-point function computed from an (A±B)/2 map.
[c]$M_{\alpha\beta}^{(\pm)}$ denotes a covariance matrix computed from an ensemble of simulated (A±B)/2 maps; separate covariance matrices were computed for the 53 and 90 GHz noise levels.

Table 4. 68% Confidence Intervals, in $\mu K^3$, for $k_2$ and $k_3$[a]

| Frequency[b] | $k_2$[c] | $k_3$[d] |
|---|---|---|
| 53 GHz | $-169 < k_2 < 644$ | $-9 < k_3 < 534$ |
| 90 GHz | $-1223 < k_2 < 113$ | $-1448 < k_3 < -353$ |
| Average | $-323 < k_2 < 371$ | $-212 < k_3 < 276$ |

[a]Approximate 68% confidence intervals, derived from the pseudo-collapsed configuration, include uncertainties due to cosmic variance.
[b]The line denoted "Average" gives weighted average confidence intervals under the assumption that the 53 and 90 GHz uncertainties are uncorrelated.
[c]$k_2$ is the coefficient of the model function $[2C_2(2.6°)C_2(\theta) + C_2^2(\theta)]$
[d]$k_3$ is the coefficient of the model function $C_2(2.6°)C_2^2(\theta)$

– 12 –

---

This manuscript was prepared with the AAS LaTeX macros v3.0.

...

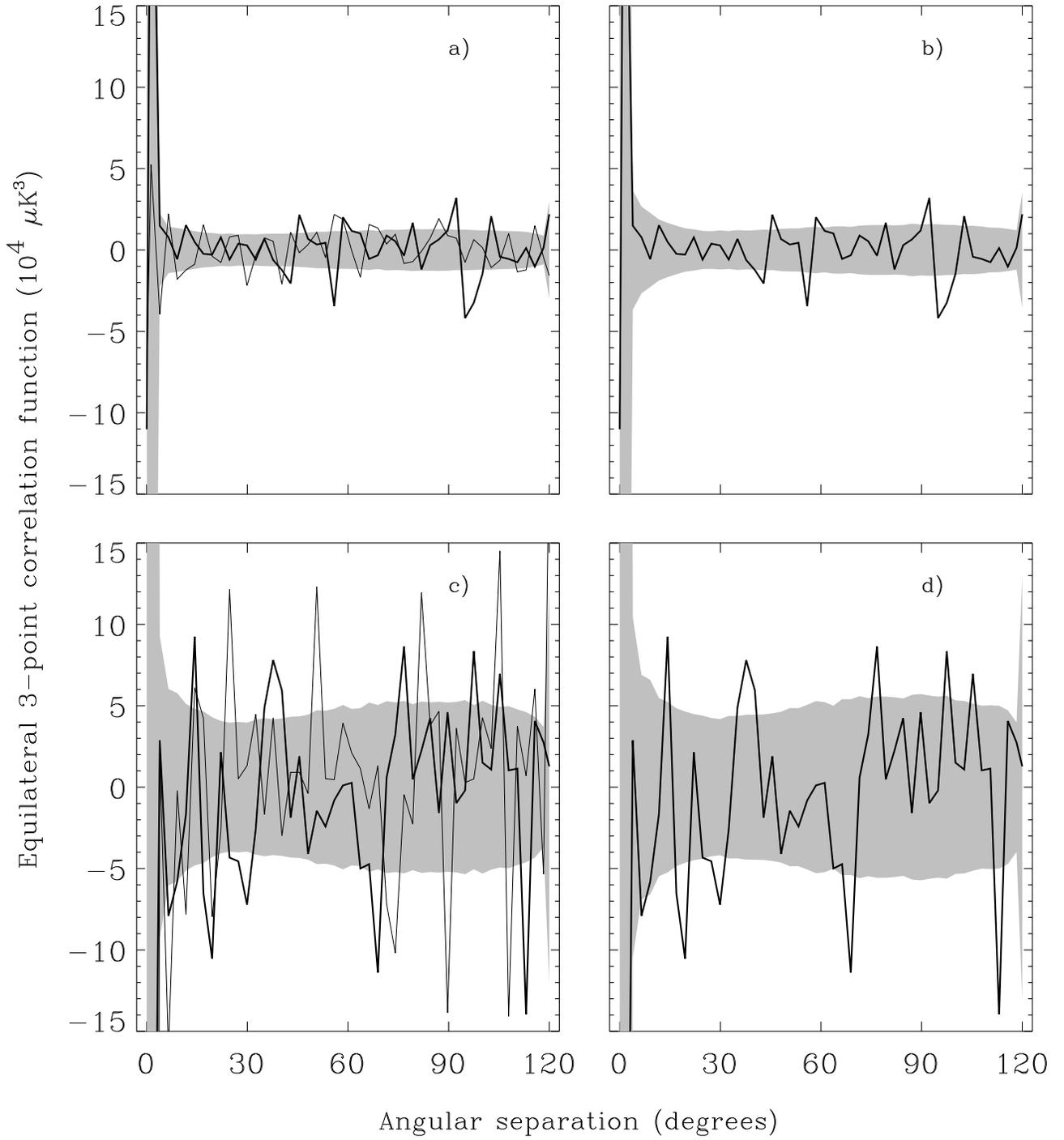

Fig. 1.— a) 53 GHz equilateral three-point correlation functions. The thick line is the result for the (A+B)/2 map, the thin line is for the (A−B)/2 map. The grey band represents the rms scatter due to instrument noise, see the text for details. b) 53 GHz equilateral three-point function for the (A+B)/2 map. The grey band represents the rms scatter due to simulated CMB sky signal plus instrument noise, see the text for details. c) 90 GHz equilateral three-point correlation functions, same format as a). d) 90 GHz equilateral three-point correlation function, same format as b).

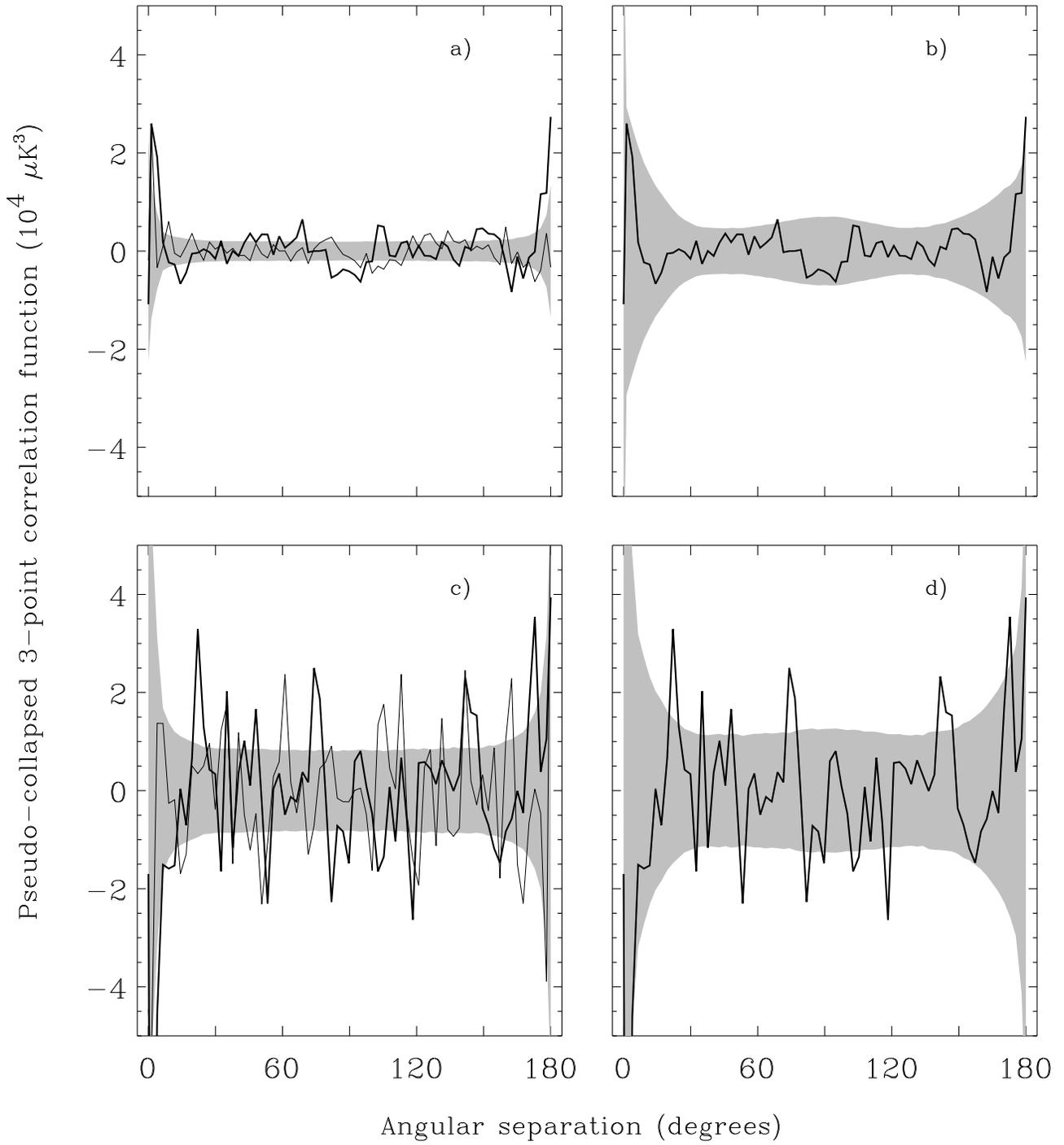

Fig. 2.— *a)* 53 GHz pseudo-collapsed three-point correlation functions, same format as Figure 1*a*. *b)* 53 GHz pseudo-collapsed three-point function, same format as Figure 1*b*. *c)* 90 GHz pseudo-collapsed three-point correlation functions, same format as Figure 1*c*. *d)* 90 GHz pseudo-collapsed three-point correlation function, same format as Figure 1*d*.